# Probing thin films and their interfaces via angle-resolved second-harmonic generation

J. Lukeš[1,2*], D.S. Gaur[1], V. Kanclíř[1], K. Žídek[1]

[1]Institute of Plasma Physics of the Czech Academy of Sciences, U Slovanky 2525/1a, 182 00 Prague 8 - Liben, Czech Republic

[2] Technical University of Liberec, Studentská 1402/2, 461 17, Liberec, Czech Republic

*Author to whom any correspondence should be addressed.

E-mail: lukesj@ipp.cas.cz

## Abstract

Increasing demands on the performance and reliability of optical thin films stimulate the development of more sensitive characterisation methods capable of investigating thin-film interfaces or local mechanical stress. Second-harmonic generation (SHG) is a phenomenon that can serve as a non-destructive probe, tracing local symmetry changes in the film surface, bulk, and buried interfaces. Here, we investigate the decomposition of the SHG response from these three regions by combining angle- and polarisation-resolved measurements. We utilised a theoretical model of SHG accounting for multiple reflections and optical interference in a thin layer. The theory was first experimentally evaluated using $Ta_2O_5$ films deposited on different glass substrates. The fitting procedure accurately reproduced the measured SHG responses, revealing pronounced substrate-dependent variations in the buried-interface contribution, while the surface and effective-bulk responses remained similar. As the next step, we studied the reliability of the SHG signal decomposition into the surface, bulk and buried-interface components. The decomposition was evaluated by generating synthetic datasets and testing whether their retrieval was unique and accurate. The influence of film thickness, measurement noise, uncertainties in linear-optical parameters, and the polarisation channels used on the retrieval of nonlinear susceptibilities was also examined. Our results demonstrate that reliable separation of SHG contributions requires accurate linear optical parameters, global fitting of multiple polarisation channels, and sufficiently pronounced interference features, outlining the capabilities and practical limitations of SHG-based region-specific characterisation of optical thin films.

## 1. Introduction

Thin films are essential components of optical, electronic, and photonic systems, serving as dielectric coatings, protective layers, waveguides, passivation films, and functional optical materials. Their performance depends not only on composition and thickness, but also on microstructure, defect concentration, residual stress, optical losses, adhesion, and temporal stability. [1] Small variations in deposition conditions can alter these properties even when the resulting films have similar refractive indices and thicknesses. [2, 3, 4] Films with similar linear optical parameters can therefore exhibit substantially different functional behaviour, which motivates the use of more sensitive characterisation methods.

Even for the simplest case of a single layer, at least three physically distinct regions can be identified: the ambient-facing surface, the film bulk, and the buried film-substrate interface. These regions form under different physical and chemical conditions and may respond differently to changes in the substrate, deposition process, post-deposition treatment, or environmental conditions. The surface is influenced by oxidation, adsorption, contamination, and structural relaxation [5]; the buried interface may contain strain, intermixing, defects, non-stoichiometric bonding, or a transition region [6]; and the film bulk reflects the internal structure produced by the deposition process, including density, structural order, residual stress, and defect distribution [7]. Non-destructive access to region-specific information would therefore be advantageous for thin-film process development.

Second-harmonic generation (SHG) is a promising tool for depth-resolved thin-film characterisation because it is contactless, non-destructive, and intrinsically sensitive to symmetry. [8] Within the electric-dipole approximation, the second-order susceptibility $\chi^{(2)}$ vanishes in an ideal centrosymmetric bulk, whereas inversion symmetry is broken at surfaces and interfaces, allowing a nonzero electric-dipole SHG response. This makes SHG suitable for probing surfaces and buried interfaces. [8, 9, 10] Deposition can, however, produce macroscopic inversion-symmetry breaking in thin films, for example, through structural, compositional, or strain gradients or preferentially oriented defects. [11, 12] In such cases, the measured electric-dipole SHG response may therefore contain contributions from the surface, film bulk, and buried interface, making separation of the three source regions nontrivial. [9, 11] Since SHG has proven to be a sensitive tool to detect laser-induced modifications of thin films [13] or defects in optical coatings [14], the ability to pinpoint the source of the changes in the layer would be of great interest.

Several approaches to modelling and separating SHG sources in thin films have been reported in the literature, but quantitative source separation often benefits from additional experimental information or constraints on the nonlinear contributions. Koopmans et al. [15] addressed the separation of interface SHG from nonlocal multipolar bulk contributions and showed that certain bulk and interface contributions are intrinsically indistinguishable, and full separation therefore requires an additional relation between bulk tensor components. Shi et al. [16] separated surface, buried-interface, and bulk EFISH contributions using reflected and transmitted SHG from multiple film thicknesses together with a prescribed spatial profile of the space-charge field. Other studies, such as those of Abtahi et al. [17] and Koskinen et al. [18], treated the nonlinear response as originating predominantly from interface or bulk contributions rather than independently retrieving all source regions. In general, the limits within which SHG can be reliably used to disentangle the surface, bulk and buried-interface contributions have not been examined and remain unclear.

In this work, we investigated in detail the reliability of the decomposition of the SHG signal into the surface, buried interface, and bulk dipole contributions. We used a theoretical model that combines two established descriptions of SHG in thin films. The surface and buried-interface contributions were calculated based on the formalism of Gielis et al. [19], which builds on interface and thin-film treatments by Mizrahi and Sipe [20] and Koopmans et al. [15] and represents the nonlinear response at the two film boundaries as polarisation sheets. The bulk contribution calculations were based on the approach of Koskinen et al. [21], in which a distributed nonlinear polarisation is represented by elementary source sheets integrated through the film. Together, these formalisms provide the basis for modelling the coherent surface, bulk, and buried-interface SHG contributions in our reflection geometry.

We first demonstrated that the presented model was able to reproduce experimental data on micrometre-thick $Ta_2O_5$ films deposited on three different substrate materials. We measured a set of four different polarisation schemes of SHG, where we varied the polarisations of the generating and SHG waves. For each of the four schemes, we measured an angle-resolved SHG intensity and fitted the data globally with the theoretical model. We attained a fit with very high fidelity with a consistent picture of individual components: the SHG contribution from the buried interface varied from sample to sample, while the bulk and surface contributions remained similar across the samples.

Subsequently, we used the verified model to generate synthetic SHG datasets analogous to those measured experimentally. By knowing the ground truth of the synthetic datasets, we were able to study the limitations of the SHG decomposition. These tests were used to examine limitations associated with film thickness, experimental noise, imprecision in layer thickness and refractive index values, and the range of available polarisation channels.

In particular, a sufficient layer thickness providing a visible interference pattern within the SHG angular dependence proved to be a crucial precondition for SHG data fitting. To verify the findings experimentally, we measured a complete SHG dataset for two ideally homogeneous layers of $Ta_2O_5$ at several independent spots. While we were able to retrieve the same SHG decomposition for the 724 nm-thick layer, the fit yielded widely differing fitted values for the 412 nm-thick layer.

Together, the synthetic and experimental analyses establish both the capabilities and the practical limitations of the proposed SHG-based source-separation approach, which need to be taken into account when interpreting the SHG datasets.

# 2. Theoretical models

## 2.1 Thin film geometry

Due to the number of various notations and conventions used in thin-film and SHG studies, we explicitly define the notation and geometry of the thin-film optics used in this article. Our notation largely follows that of Koskinen et al. [21]. The geometry is shown schematically in Fig. 1.

The sample is represented by a single thin layer of thickness *d* on a semi-infinite substrate. We assume that both the layer and substrate are non-absorbing and that the measurements are performed sufficiently far from optical resonances at both the fundamental and SHG wavelengths, allowing the refractive indices and $\chi^{(2)}$ tensors to be treated as real-valued. [22]

SHG may originate at the ambient-film surface, throughout the film bulk, and at the buried film-substrate interface. Reflections from the rear side of the substrate are neglected. In the experiments, wedged substrates deflected these reflections away from the collection path.

The z-axis is normal to the interfaces and directed toward the ambient medium (air); the x-axis lies in the plane of incidence, and the y-axis is perpendicular to the plane of incidence. The fundamental beam is incident from the ambient side at an angle $\theta$ in the xz-plane. Within the film, $k^-$ and $k^+$ denote waves propagating toward the substrate and toward the ambient medium, respectively. The corresponding p-polarisation vectors lie in the xz-plane and are perpendicular to their wave vectors, while the common s-polarisation vector is directed along the y-axis – see Fig. 1.

Reflections at the film boundaries produce forward- and backward-propagating fundamental fields whose coherent superposition determines the local driving field in each source region. The reflected SHG field is then obtained by coherently summing the generated contributions after their propagation and multiple reflections within the layered structure. [15, 19, 23]

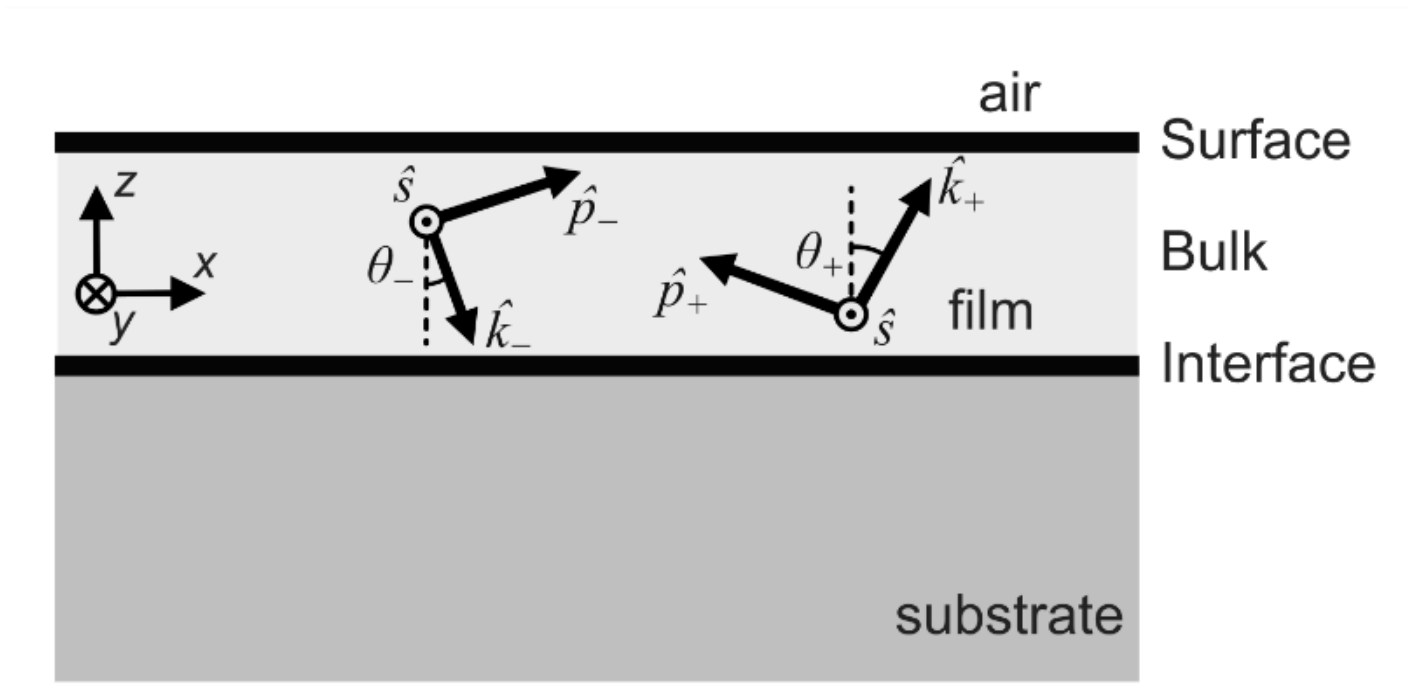

Fig. 1 - Thin-film geometry and notation used in the SHG calculations - $\hat{k}$ denotes the normalised wave vector; the − and + directions correspond to propagation toward the substrate and toward the ambient medium, respectively. Polarisation vectors are denoted by $\hat{p}$ and $\hat{s}$, and $\theta$ represents the propagation angle.

## 2.2 SHG theory

Within the electric-dipole approximation, the nonlinear polarisation at twice the fundamental frequency is [24]

$$\vec{P}^{(2)}(2\omega, z) = \epsilon_0 \boldsymbol{\chi}^{(2)}(z) : \vec{E}(\omega, z)\vec{E}(\omega, z), \quad (1)$$

where $\varepsilon_0$ is the vacuum permittivity, $E(\omega,z)$ is the local fundamental field, and $\chi^{(2)}$ is the second-order susceptibility tensor. The double contraction denotes the coupling of two fundamental-field components to each component of the nonlinear polarisation.

For an ideal centrosymmetric bulk material, all electric-dipole components of $\chi^{(2)}$ vanish due to internal symmetry restrictions. [24] At surfaces and interfaces, inversion symmetry is broken, and a nonzero electric-dipole response is allowed. [20] The effective bulk term considered here represents a layer volume with macroscopic inversion-symmetry breaking along the z-axis. Its microscopic origin is not specified by the model; possible origins include structural or compositional gradients [25], strain gradients [11], or preferentially oriented defects [26].

For the geometry described above, the spatial distribution of nonlinear sources is written as

$$\boldsymbol{\chi}^{(2)}(z) = \boldsymbol{\chi}_S^{(2)}\delta(z) + \boldsymbol{\chi}_I^{(2)}\delta(z+d) + \boldsymbol{\chi}_B^{(2)}H(-z)H(z+d) \quad (2)$$

The surface is located at $z = 0$ and the buried interface at $z = -d$. The Dirac delta functions represent the two boundaries as nonlinear polarisation sheets, while the Heaviside functions restrict the effective bulk response to $-d < z < 0$. Thus, $\boldsymbol{\chi}_S^{(2)}$ and $\boldsymbol{\chi}_I^{(2)}$ are effective sheet susceptibilities and $\boldsymbol{\chi}_B^{(2)}$ is a bulk susceptibility. Note that $\chi^{(2)}$ values from interfaces and bulk, although denoted identically in the literature, have different physical dimensions corresponding to $m^2/V$ [20] and m/V, respectively [24].

The surface and buried-interface fields are calculated using the thin-film formalism of Gielis et al. [19]. The local fundamental field at each sheet contains the incident field and the partial waves generated by repeated reflections at both film boundaries. The nonlinear polarisation is therefore driven by a polarisation- and angle-dependent local field that can differ substantially from the incident field. The emitted second-harmonic field is propagated toward the ambient medium with the appropriate Fresnel coefficients, phase factors, and multiple reflections at $2\omega$.

While the detailed derivation of the theory can be found elsewhere [15, 19], here we provide the expression implemented in our code:

$$E_L(2\omega) = \frac{\mathrm{i}\omega}{c\cos(\theta)} \vec{e}_{out} \left(\sum_L e_L(2\omega) \cdot \boldsymbol{\chi}_L^{(2)} : \overrightarrow{E_L}(\omega)\overrightarrow{E_L}(\omega)\right) E_0^2(\omega), \quad (3)$$

where $L \in S, I$, $e_{out}$ is the normalised polarisation vector of the detected reflected SH field (in our case, either $p$ or $s$ polarisation), $e_L(E_L)$ contains Fresnel factors and multiple-reflection terms at the SH (fundamental) frequency and $E_0$ is the amplitude of the incident fundamental field.

The bulk contribution is evaluated using the Green-function approach used by Koskinen et al. [21]. The film is treated as a continuous distribution of elementary nonlinear source sheets. The field generated at each depth is propagated to the detector and integrated coherently over the film thickness. Both the driving field and the propagation factor depend on position within the layer, so sources at different depths contribute with different amplitudes and phases. The bulk response depends on phase mismatch and the cavity response of the stack and is not, in general, proportional to film thickness alone. In the present work, the formalism is evaluated for the reflected output field.

Analogously to the previous case, the detailed derivation of the theory can be found elsewhere [21], and here we provide the expression implemented in our code:

$$E_B(2\omega) = \frac{\mathrm{i}(2\omega)^2}{2c^2 k_z} \vec{e}_{out} \boldsymbol{T}_{21}^{2\omega} \boldsymbol{C}^{2\omega} \Big( \sum_{\beta,\gamma} \boldsymbol{\chi}_B^{(2)} : \overrightarrow{E_\beta}(\omega) \overrightarrow{E_\gamma}(\omega) \Phi_{+\beta\gamma} + \mathrm{e}^{\mathrm{i}2k_z D} \boldsymbol{R}_{23}^{2\omega} \sum_{\beta,\gamma} \boldsymbol{\chi}_B^{(2)} : \overrightarrow{E_\beta}(\omega) \overrightarrow{E_\gamma}(\omega) \Phi_{-\beta\gamma} \Big), \tag{4}$$

where $k_z$ is the $z$ component of the wave vector of the generated SH wave, $e_{out}$ is the normalised polarisation vector of the detected reflected SH field, $\boldsymbol{T}$ ($\boldsymbol{R}$) is the Fresnel transmission (reflection) tensor, $\boldsymbol{C}$ denotes the multiple-reflection factor for the SH field inside the medium, $\beta, \gamma \in +, -$ denote the propagation directions and $\Phi_{\alpha\beta\gamma}$ is the phase-mismatch factor between the fundamental and SH fields.

The numerical implementation was cross-checked against a transfer-matrix method (TMM) model using identical optical and nonlinear inputs, yielding consistent reflected-SHG results. A second validation compared the Gielis boundary-source formalism with the Koskinen Green-function formulation in the limiting case of a single nonlinear interface between two media. The source was represented directly as an interface polarisation sheet in the Gielis model and as an ultrathin nonlinear layer with the same integrated dipole response in the Koskinen model.

The reflected SHG signal is obtained from the coherent sum of the surface, bulk, and buried-interface fields after projection onto the selected output polarisation. The measured intensity is proportional to the squared modulus of this total field. Intensities calculated for individual source regions, therefore, cannot generally be added, because cross terms may interfere constructively or destructively. This interference is important for fitting: a weak isolated contribution may still modify the total response through interference with the other fields.

The three source regions were treated as in-plane isotropic and described by $C_{\infty v}$ symmetry. With permutation symmetry in the two fundamental-field indices, the three independent components are [27]

$$\chi^{(2)}_{zzz}, \qquad \chi^{(2)}_{zxx} = \chi^{(2)}_{zyy}, \qquad \chi^{(2)}_{xxz} = \chi^{(2)}_{xzx} = \chi^{(2)}_{yyz} = \chi^{(2)}_{yzy}$$

Different polarisation configurations are sensitive to different combinations of independent tensor components. Here we denote these configurations as pP, sP, mP and mS, where the first lowercase letter denotes the input polarisation of the fundamental field and the second uppercase letter denotes the polarisation of the detected SH-field, and “m” corresponds to mixed input polarisation with equal *p*- and *s*-polarised components. The pP polarisation can contain all three independent tensor components and therefore carries a large amount of information, but it also permits correlations among several tensor components and source regions. Within the assumed $C_{\infty v}$ symmetry, the sP configuration isolates $\chi^{(2)}_{zxx}$, and mS primarily constrains $\chi^{(2)}_{xxz}$. The mP configuration provides an additional coherent combination excited by equal *p*- and *s*-polarised fundamental components. [19] Simultaneous analysis of pP, sP, mP, and mS therefore provides complementary constraints that are not available from a single polarisation channel.

The presented model is restricted to electric-dipole SHG in linearly isotropic media. We neglected electric-field-induced SHG, electric-quadrupole and magnetic-dipole contributions, and nonlinear radiation from the substrate and rear-surface reflection.

## 3. Experimental details

### 3.1 Samples and deposition

The $Ta_2O_5$ films studied in this work were deposited by ion-assisted electron-beam physical vapor deposition. The nominal chamber temperature was 160 °C, and the ion source was operated at an acceleration voltage of 250 V and a source current of 1.4 A.

The deposited samples were $Ta_2O_5$ films with thicknesses ranging from 400 nm to 3000 nm. Wedged N-BK7, N-SF10 and fused silica were used as substrates for these films. The wedge on the backside of the substrates deflected rear-surface reflections outside the collection path during SHG measurements.

Film thicknesses and refractive indices were initially determined by spectroscopic ellipsometry. Because the calculated SHG interference structure is highly sensitive to thickness, the local film thickness at the measured spot was refined from the angular dependence of the reflected fundamental beam, which was recorded simultaneously with the SHG signal. The refined thickness was then fixed during SHG fitting. This accounts for local thickness variations caused by deposition inhomogeneity and reduces the risk that errors in the linear-optical input are compensated by changes in the fitted nonlinear susceptibilities.

### 3.2 Measurement setup

Angle-resolved SHG measurements were performed in reflection geometry using a femtosecond Yb:YAG Pharos laser (Light Conversion) at 1028 nm, corresponding to an SHG wavelength of 514 nm. The repetition rate was 100 kHz, the pulse duration was approximately 225 fs, and the beam diameter at the sample was approximately 25 μm.

The average incident power was adjusted between 40 and 44 mW using a half-wave plate and a linear polariser and was held constant during each scan. This range corresponds to pulse energies of approximately 0.40–0.44 μJ/pulse.

The sample was mounted on a motorised rotation stage and measured from 30° to 70° incidence in 0.5° steps. At each angle, the pP, sP, mP, and mS responses were recorded. In addition to these four symmetry-allowed configurations, symmetry-forbidden channels were measured as controls and produced signals at approximately the experimental noise level.

After reflection, the combined fundamental and second-harmonic beam passed through the output analyser and was then separated spectrally into two detection channels. The SHG component was further isolated using spectral filters and detected with a photomultiplier tube, while the reflected fundamental signal was measured in parallel. Both signals were digitised via a data-acquisition card and processed by a software implementation of a lock-in amplifier. A detailed description of the optical arrangement is given in [28]. Due to the lack of a calibration measurement, the fitted susceptibilities $\chi^{(2)}$ are reported in relative units.

## 4. Results and discussion

### 4.1 SHG in a thick $Ta_2O_5$ film

Figure 2 shows angle-resolved SHG measured from a 2885 nm-thick $Ta_2O_5$ film on an N-BK7 substrate in the pP, sP, mP, and mS configurations. The four scans were acquired sequentially and fitted simultaneously as a global fit using a single set of surface, buried-interface, and effective-bulk susceptibility components $\chi^{(2)}$. The large layer thickness produces pronounced angular SHG modulation, including several maxima and minima in pP and mP and distinct structure in sP and mS. These features arise from the interference of the fundamental fields within the film and coherent interference of the SHG fields generated in the spatially separated source regions.

The complete surface–interface–bulk model reproduced the main interference features in all four polarisation channels of the SHG datasets, including their angular positions and relative amplitudes. It is worth noting that the model retained the same amplitude scaling throughout all the polarisation configurations. We verified that fits with various random sets of initial parameters that converged to good agreement between experiment and theory led to the same solution presented in Figure 2.

In contrast, when the model is restricted to either the effective-bulk source alone or the surface and buried-interface sources alone, the calculated curves fail to reproduce several features of the measured angular response and yield substantially larger residuals. Since the disagreement occurs across multiple polarisation configurations, it is unlikely to arise from a channel-specific experimental artefact. Within the assumed electric-dipole model, the data therefore require contributions from both boundary and effective-bulk source terms. The simultaneous agreement across all four polarisation channels provides an additional constraint on this decomposition because the individual configurations probe different combinations of susceptibility components. [19]

Figure 2 also shows, in each panel, the isolated contribution of the bulk SHG (blue dashed line) and boundary SHG (dotted green line). With the exception of the sP configuration, the bulk and boundary SHG responses have comparable magnitudes over most parts of the angular range and are therefore non-negligible. These isolated intensity curves should not be interpreted as additive components of the complete response, because the corresponding complex fields interfere coherently.

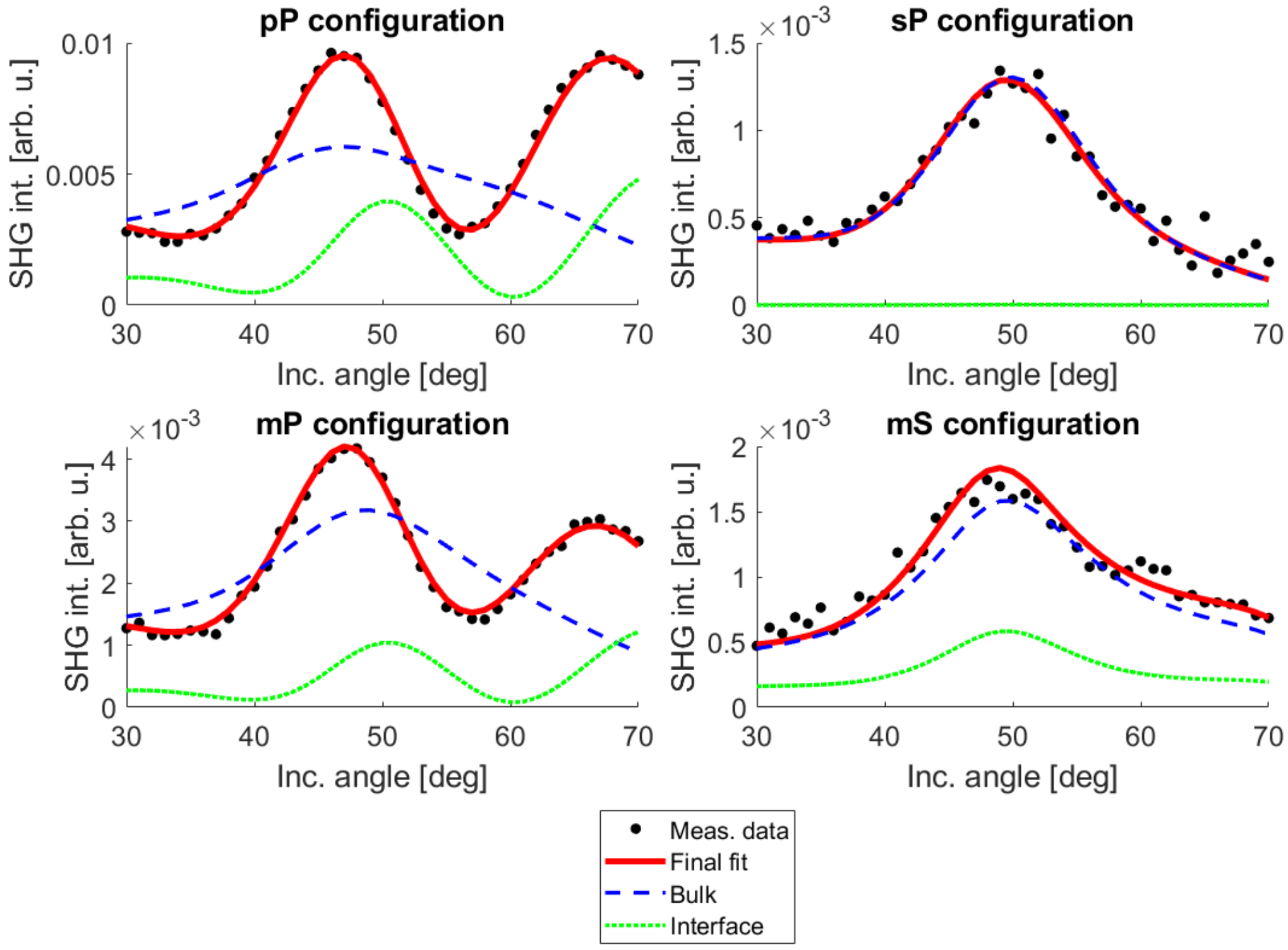

Fig. 2 – **Black dots** – measured data for angle-dependent reflected SH from a 2885 nm-thick $Ta_2O_5$ film deposited on an N-BK7 substrate for four different polarisation configurations (the first lowercase letter denotes the input polarisation, the second uppercase letter denotes the detected polarisation, and "m" corresponds to mixed p/s polarisation); **Red solid line** – global fit using the model described in the SHG theory section; **Blue dashed line** – isolated bulk contribution; **Green dotted line** – combined surface and buried-interface contribution.

## 4.2 Substrate dependence of the buried-interface response

For further experimental testing of the model, approximately 2885 nm-thick $Ta_2O_5$ films were co-deposited on N-BK7, N-SF10, and fused-silica substrates, which are denoted here as the N-BK7, N-SF10, and FS samples. Ellipsometry measurements yielded the same refractive-index dispersion and thicknesses across all samples. This was confirmed by fitting thicknesses from measured reflected fundamental radiation, which showed a subtle difference in thicknesses of less than 0.5%. The $Ta_2O_5$ film bulks and ambient-facing surfaces were therefore expected to be closely comparable, whereas the buried $Ta_2O_5$/substrate interfaces were expected to vary depending on the substrate type.

Figure 3 shows the measured and fitted responses for all three samples. Pronounced interference is present in each sample, with substrate-dependent differences in amplitude and angular dependence. Over much of the angular range, the FS sample generated more than twice the SHG intensity of the N-BK7 sample signal and approximately four times that of the N-SF10 sample. In the pP configuration, the N-BK7 and N-SF10 samples show maxima near 45° and 68°, but their relative amplitudes differ: the maxima are comparable for the N-BK7 sample, whereas the lower-angle maximum is weaker for the N-SF10 sample. The FS sample likewise shows two maxima of comparable amplitude, shifted to approximately 43° and 63°. The opposite trend appears in the mP configuration, where, for the FS and N-BK7 samples, one of the maxima is lower than the other. The shape and magnitude of the sP and mS responses also differ among the samples.

These differences cannot be assigned directly to nonlinear susceptibilities because the substrate refractive index also changes the Fresnel coefficients, local fundamental fields, SHG propagation, and out-coupling. Each sample was therefore fitted using its own linear-optical parameters and the complete nonlinear model, which reproduces the characteristic features of all four polarisation configurations.

Table 1 summarises the fitted relative susceptibility components. The fitted surface and effective-bulk components are substantially more similar among the three samples than the buried-interface components, which vary strongly with substrate. This agrees with the sample design: the co-deposited $Ta_2O_5$ bulks and ambient-facing surfaces are expected to be comparable, whereas the substrate-facing interface is the intentionally modified source region. The agreement between the sample design and the fitted source-specific response provides an experimental consistency test of the decomposition and indicates that the fit retains sensitivity to source location. It also makes an explanation based solely on redistribution among correlated parameters less likely.

The same data scaling factor was used for all three samples, allowing direct comparison of a given fitted component between samples. However, the sheet susceptibilities of the surface and buried interface should not be compared numerically with the bulk susceptibility because the sheet and bulk quantities have different physical dimensions.

Finally, it is worth stressing that this comparison was made possible by choosing three analogous substrates from optical glasses. In general, substrate-dependent nucleation and growth may alter residual stress, defect structure, density, or morphology in addition to changing the chemical interface, and some tensor components remain weakly constrained.

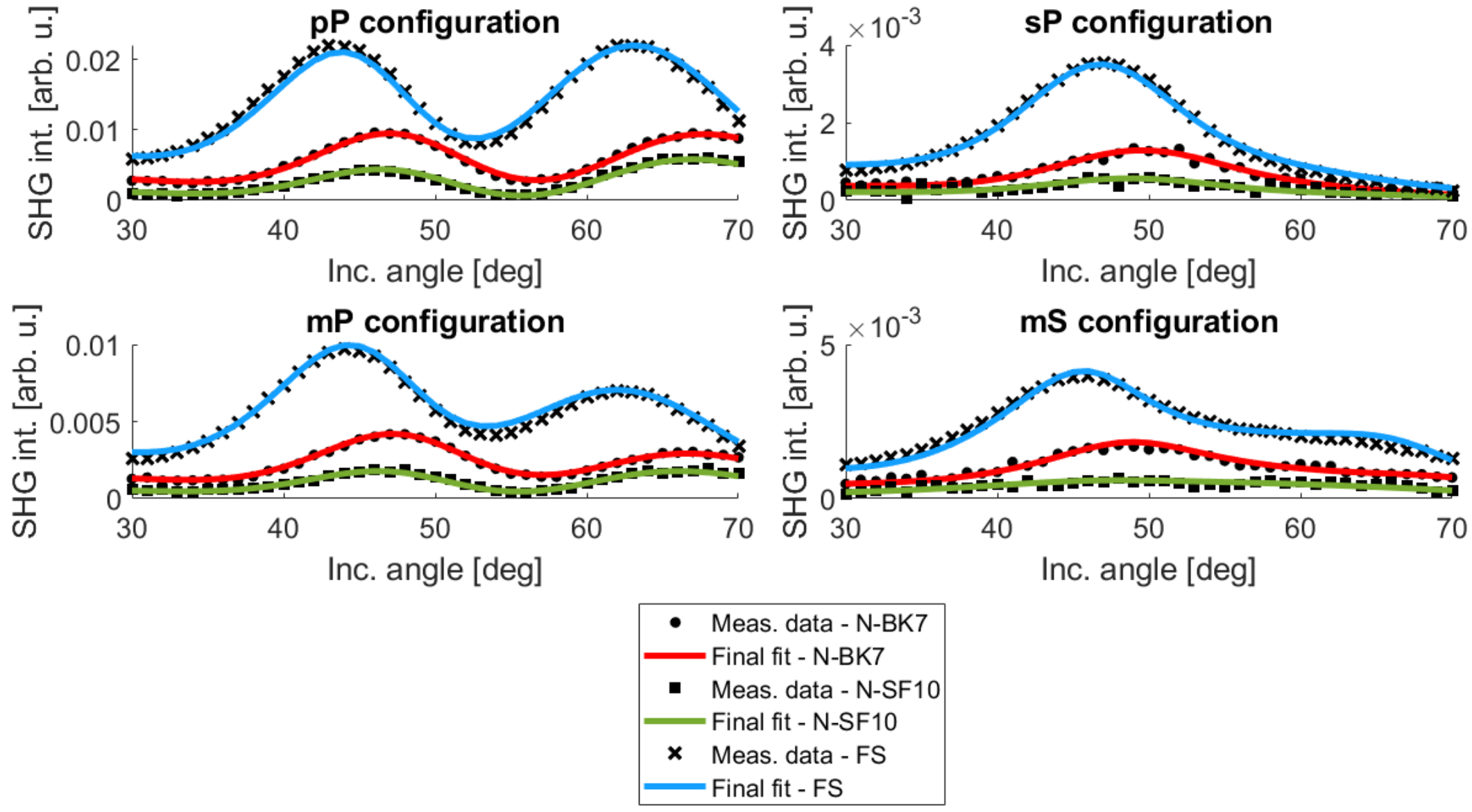

Fig. 3 – Angle-dependent measured SH data in four polarisation configurations and corresponding fitted curves using the full nonlinear model for approximately 2.9 µm-thick $Ta_2O_5$ films co-deposited on N-BK7, N-SF10, and fused-silica (FS) substrates. The first lowercase letter in a polarisation configuration denotes the input polarisation, the second uppercase letter the detected polarisation, and "m" denotes mixed *p*/*s* input polarisation.

| | $\chi^{(2)}$ Surface [arb. u.] | | | $\chi^{(2)}$ Interface [arb. u.] | | | $\chi^{(2)}$ Bulk [arb. u.] | | |
|---|---|---|---|---|---|---|---|---|---|
| | $\chi_{zzz}$ | $\chi_{zxx}$ | $\chi_{xxz}$ | $\chi_{zzz}$ | $\chi_{zxx}$ | $\chi_{xxz}$ | $\chi_{zzz}$ | $\chi_{zxx}$ | $\chi_{xxz}$ |
| **N-BK7** | -27 | 0 | 0.3 | 10 | 0 | -3.9 | 0.2 | 0.2 | 0.2 |
| **N-SF10** | -25 | -0.5 | 0.3 | 1.9 | 3.0 | -0.6 | 0.8 | 0.1 | 0.2 |
| **FS** | -19 | -0.1 | 0.7 | 83 | -0.8 | 1.1 | 0.2 | 0.2 | 0.2 |

Tab. 1 – Fitted $\chi^{(2)}$ values corresponding to the curves in Fig. 3. Note that values are not absolutely calibrated, and surface and buried-interface susceptibilities cannot be compared numerically with the bulk susceptibility because they have different physical dimensions ($m^2/V$ for surface/interface terms and m/V for the bulk term). Each source region is represented by one $\chi^{(2)}$ tensor with three independent nonzero components under the assumed symmetry.

## 4.3 SHG decomposition limitations studied via synthetic $Ta_2O_5$ data

In the previous section, the model was tested on the experimental datasets for optically thick films, where pronounced interference structure provides strong constraints on the fit. To assess the reliability of susceptibility retrieval for thinner films and for incomplete datasets, we performed a systematic analysis using synthetic data. Four main findings are discussed below.

The forward model was used to generate synthetic datasets, which were then repeatedly fitted from many randomised initial parameters. We considered parameter recovery reliable when fits with different initial values reproduced the synthetic data and converged to the generating $\chi^{(2)}$ parameters. In poorly identifiable cases, however, substantially different $\chi^{(2)}$ parameter sets reproduced the same dataset with comparably small residuals. This procedure tests the stability of parameter recovery rather than the model's physical completeness. Unless stated otherwise, the tests used synthetic data for a 600 nm $Ta_2O_5$ film on an N-BK7 substrate. The data included nonzero surface, buried-interface, and bulk contributions and were perturbed with intensity-dependent Gaussian noise adjusted to reproduce the overall appearance of the experimentally measured SHG curves. The known generating $\chi^{(2)}$ parameters provided a controlled test of the recovery procedure.

Because the same forward model generated and fitted the synthetic data, these tests evaluate identifiability within the assumed model rather than its physical completeness. They nevertheless show that small residuals are necessary but not sufficient for reliable susceptibility recovery.

### Fitting of multiple polarisation schemes

Figure 4(a) shows the synthetic pP data fitted with a reduced model containing only surface and buried-interface SHG (green dotted line). The fit with a reduced model is misleading in this case, because the synthetic data themselves were generated with nonzero surface, buried-interface, and bulk contributions. Nevertheless, despite omitting the bulk term, the reduced model yields small residuals across the full pP angular range. The same ambiguity was also observed for thicker films in our tests. Therefore, a pP-only fit by itself cannot be used to identify the physical source of the SHG signal, despite being in perfect agreement with the data.

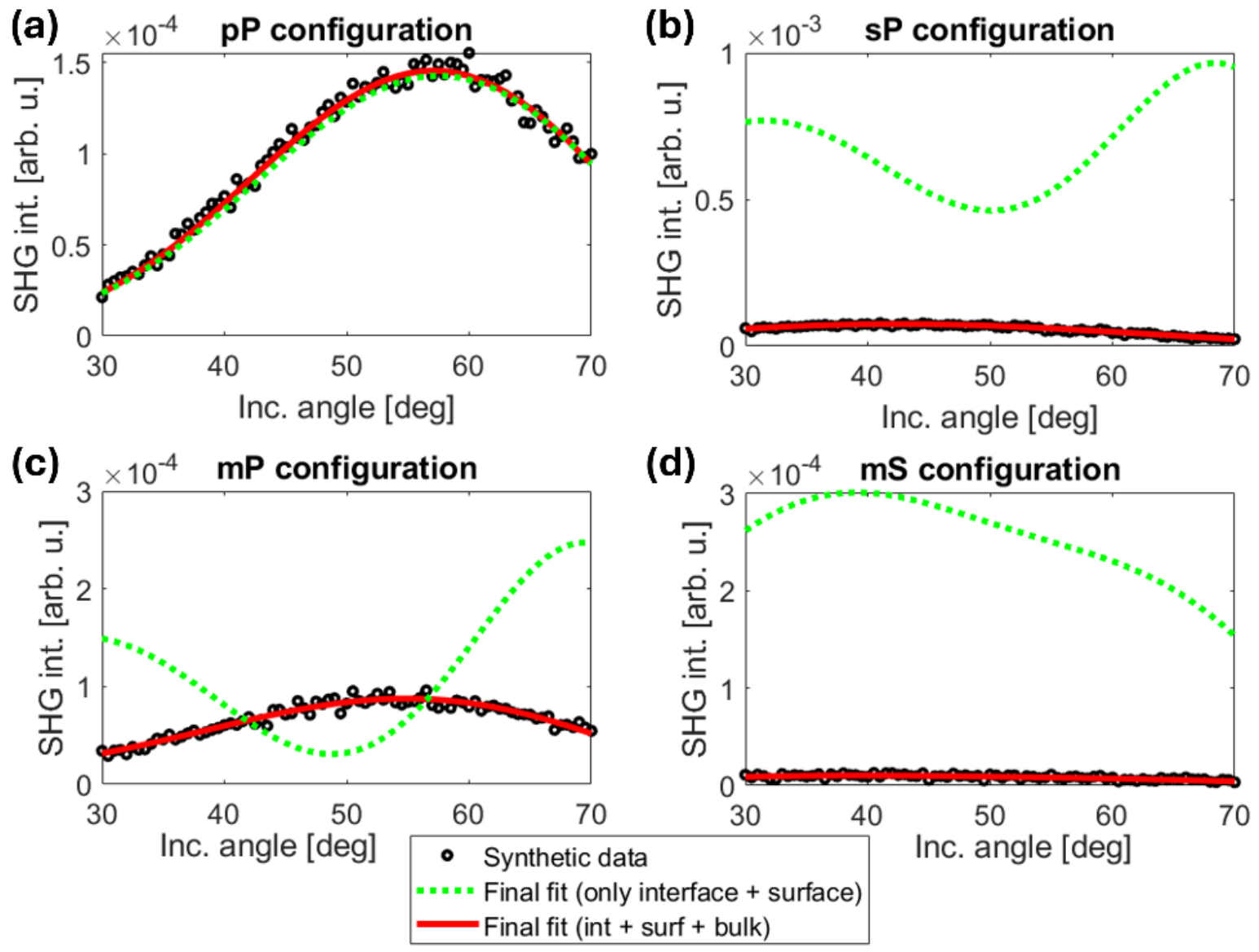

Fig. 4 – **Illustration of the need for multiple polarisation configurations. Black circles** – synthetic angle-dependent SHG data for a 600 nm $Ta_2O_5$ film on an N-BK7 substrate, with all components non-zero (surface + interface + bulk) and with added random noise; **Green dotted curve** – fit with bulk $\chi^{(2)}$ values fixed to zero, for which only the pP configuration (**panel (a)**) was used for fitting; the rest of the plotted green curves were calculated from these fitted parameters; **Red solid curve** – fit with all components allowed.

To illustrate the importance of including all polarisation configurations in the fitting procedure, we used the surface and buried-interface $\chi^{(2)}$ values obtained from the reduced pP-only fit to calculate the expected sP, mP and mS responses, shown by the green dotted curves in Fig. 4(b-d). Although the pP response is reproduced well, the predictions for sP, mP, and mS show clear discrepancies in both the magnitude and the angular dependence of the SH response.

On the contrary, it is possible to find a simultaneous fit of all four polarisation configurations with the complete model, which produces good agreement with the synthetic data, as shown by the red solid curves in Fig. 4(a-d). The recovered parameters for the full fit also agree well with the generating values (see Table 2), indicating that even for a 600 nm-thick film, the $\chi^{(2)}$ components can be reliably recovered under the tested model and noise conditions.

| | **$\chi^{(2)}$ Surface [arb. u.]** | | | **$\chi^{(2)}$ Interface [arb. u.]** | | | **$\chi^{(2)}$ Bulk [arb. u.]** | | |
|---|---|---|---|---|---|---|---|---|---|
| | $\chi_{zzz}$ | $\chi_{zxx}$ | $\chi_{xxz}$ | $\chi_{zzz}$ | $\chi_{zxx}$ | $\chi_{xxz}$ | $\chi_{zzz}$ | $\chi_{zxx}$ | $\chi_{xxz}$ |
| **Surface + Interf.** | -2.6 | 3 | 0.8 | 0.9 | 5.8 | 2.7 | 0 | 0 | 0 |
| **Full model** | 0.6 | 0.2 | 0.6 | -0.5 | 0.7 | 0.3 | -0.01 | -0.1 | -0.04 |
| **Ground Truth** | 0.9 | 0.2 | 0.6 | 0.4 | 0.8 | 0.3 | 0.03 | -0.1 | -0.04 |

Tab. 2 – Comparison of fitted parameters for the 600 nm $Ta_2O_5$ synthetic dataset on N-BK7. First row: reduced surface-plus-interface model fitted using only the pP configuration. Second row: complete model fitted simultaneously using all four polarisation configurations. Third row: ground-truth $\chi^{(2)}$ values used to generate the synthetic data. The corresponding data and fitted curves are shown in Fig. 4.

**Conclusion I. Multiple polarisation configurations are required.** A small residual in one polarisation channel does not establish either correct susceptibility values or correct source attribution. In pP, several tensor components and source regions enter the same signal, producing strong parameter correlations. Simultaneous fitting of pP, sP, mP, and mS restricts the susceptibility combinations consistent with the complete dataset and improves source identifiability. The additional channels should therefore be treated as essential constraints rather than merely as supplementary measurements.

### Role of uncertainty in linear optical parameters

We next examined how uncertainty in the linear-optical parameters propagates into the recovered nonlinear susceptibilities. Synthetic datasets were generated with systematically shifted refractive indices, while the film thickness was changed simultaneously to preserve the optical path length. This correlated variation was chosen to represent the coupling between refractive index and thickness that can occur in ellipsometric analysis [29]. During SHG fitting, the refractive index was kept at its nominal value, whereas the thickness was re-estimated from the reflected fundamental signal, following the procedure used for the experimental data. Figure 5 shows the resulting median relative deviation of the fitted $\chi^{(2)}$ values from the ground truth for three tested film thicknesses.

As shown in Fig. 5, the retrieval error is smallest near zero refractive-index mismatch, as expected, and generally increases as the mismatch grows. Correlated errors in refractive index and thickness propagate into the extracted nonlinear susceptibilities even when the film thickness is independently refined from the reflected fundamental signal. The magnitude and shape of this effect differ among the three tested film thicknesses, indicating that the sensitivity depends on the specific thin-film configuration and laser wavelength rather than following a simple universal trend.

**Conclusion II. Accurate linear-optical parameters are essential.** A refractive-index mismatch below 1% is generally sufficient to ensure a reliable fit.

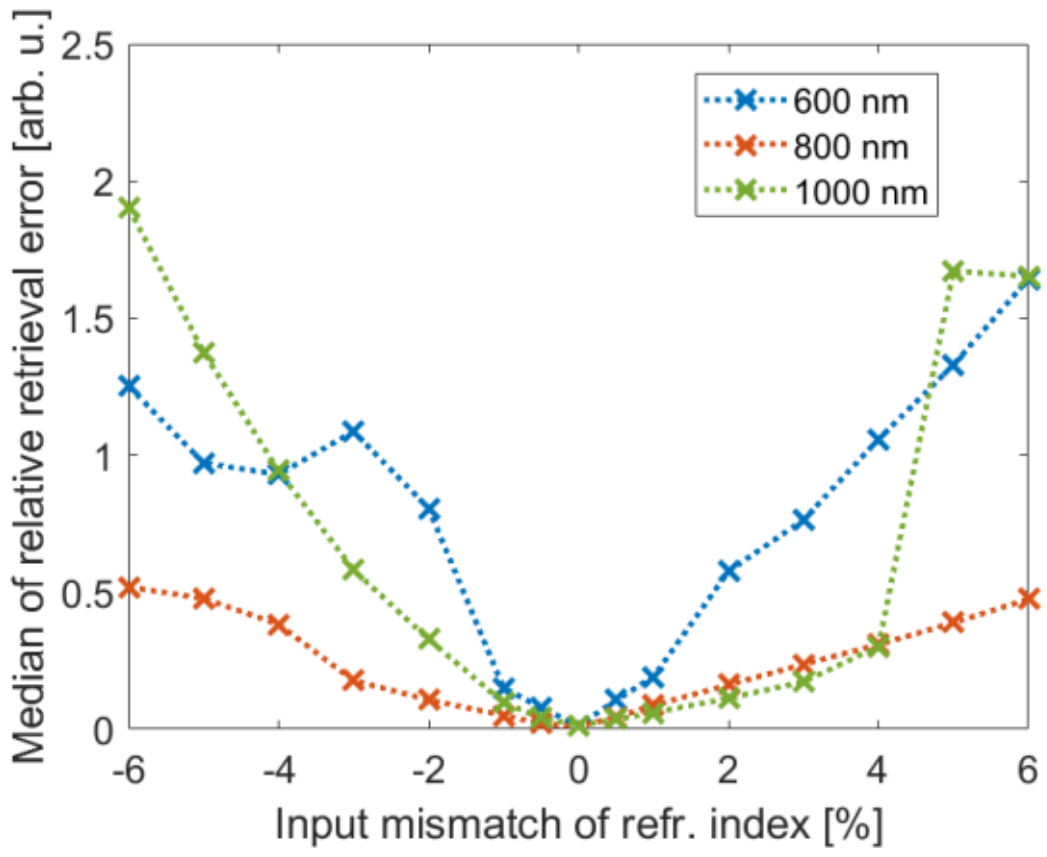

Fig. 5. Relative retrieval error for synthetic $Ta_2O_5$ datasets of three film thicknesses on N-BK7 as a function of refractive-index mismatch. The refractive index used to generate each dataset was intentionally altered, and the generating thickness was adjusted to preserve the original optical path length. During fitting, the nominal refractive index was used and the thickness was re-estimated from the reflected fundamental signal.

**Uncertainty of individual components**

In addition to the polarisation configuration, the angle of incidence helps distinguish individual $\chi^{(2)}$ components by changing the projections of the p-polarised electric field onto the x- and z-axes. In high-index materials such as $Ta_2O_5$, refraction reduces the propagation angle inside the layer and thus the z-component of the p-polarised field. The $\chi^{(2)}_{zzz}$ component, therefore, produces only small changes in the calculated intensity. As a result, the SHG signal is relatively insensitive to the $\chi^{(2)}_{zzz}$ components.

This reduced sensitivity is also evident in Table 3, which summarises fits of 30 synthetic datasets generated with identical input parameters but independent random-noise realisations. We used a higher noise level than observed experimentally to make differences in parameter sensitivity more apparent. In all three source regions, $\chi^{(2)}_{zzz}$ exhibits a larger spread of fitted values than $\chi^{(2)}_{zxx}$ and $\chi^{(2)}_{xxz}$, as quantified by the standard deviation (STD). The relative standard deviation (RSD) shows the same trend, although it becomes poorly conditioned when the mean fitted value approaches zero, as for the bulk $\chi^{(2)}_{zzz}$ component. This limited sensitivity is a property of the present optical geometry rather than a numerical failure of the minimisation. Values of $\chi^{(2)}_{zzz}$ should therefore be interpreted with particular caution.

**Conclusion III. The $\chi^{(2)}$zzz components are comparatively weakly constrained.**

| | $\chi^{(2)}$ Surface [arb. u.] | | | $\chi^{(2)}$ Interface [arb. u.] | | | $\chi^{(2)}$ Bulk [arb. u.] | | |
|---|---|---|---|---|---|---|---|---|---|
| | $\chi_{zzz}$ | $\chi_{zxx}$ | $\chi_{xxz}$ | $\chi_{zzz}$ | $\chi_{zxx}$ | $\chi_{xxz}$ | $\chi_{zzz}$ | $\chi_{zxx}$ | $\chi_{xxz}$ |

| Fitted values | -10 | 5 | -10 | -25 | 24 | 28 | 0.2 | -1.0 | -1.3 |
|---|---|---|---|---|---|---|---|---|---|
| **STD** | 14 | 2 | 4 | 31 | 3 | 11 | 12 | 0.3 | 0.5 |
| **RSD** | 1.4 | 0.5 | 0.4 | 1.3 | 0.1 | 0.4 | 62 | 0.3 | 0.4 |
| **Ground truth** | -10 | 5 | -11 | -20 | 25 | 30 | -1.6 | -1 | -1.4 |

Tab. 3 – Mean fitted values, standard deviation (STD), and relative standard deviation (RSD) for 30 synthetic datasets representing the same 600 nm $Ta_2O_5$ film on N-BK7. Independent noise realisations were added to each dataset. The imposed noise level was intentionally higher than that typically observed experimentally to make differences in parameter sensitivity more apparent.

## 4.4 Layer thickness limitation

We carried out a set of synthetic data fitting tests, which indicated that under the present measurement and model conditions, parameter recovery becomes more stable for $Ta_2O_5$ film thicknesses above approximately 500 nm. We therefore tested experimental reproducibility on two $Ta_2O_5$ films deposited on N-BK7, with thicknesses of 412 and 724 nm. The samples were chosen to have high lateral thickness uniformity so that neighbouring measurement positions were expected to exhibit identical angular responses and fitted source contributions.

First, we measured the 724 nm $Ta_2O_5$ film at two neighbouring positions separated by about 1 mm. The four polarisation configurations were fitted independently at each position using the complete model and the locally determined thicknesses. Both datasets were reproduced with comparably small residuals, and the two fits converged to closely similar susceptibility components despite independent measurement noise. An analogous experiment was carried out for the 412 nm $Ta_2O_5$ thin film deposited on an N-BK7 substrate. As in the previous experiment, both datasets were reproduced with small residuals, but the fitted susceptibility components differed substantially across positions.

As a result, our experimental data confirmed that, for the given experimental configuration, a reliable SHG angle-dependence analysis is possible only for samples with optical path lengths greater than the wavelength of the generated wave, and that the angle dependence leads to pronounced interference patterns. However, note that the actual thickness limit depends on the laser wavelength, angular range, refractive-index contrast, polarisation set, and noise level. The thickness limit is therefore specific to the investigated configuration and should not be interpreted as a universal limit for thin-film SHG analysis.

Although the comparison does not prove mathematical uniqueness of the tensor components and assumes that the film is laterally homogeneous over the measured separation, the thickness variation between positions separated by several millimetres is expected to be below 0.1%, which, based on the preceding robustness analysis, lies within the range that does not significantly affect the fitting procedure. The comparison, therefore, provides an experimental consistency test of the fitting procedure.

**Conclusion IV. Parameter recovery becomes unstable when the angular response lacks pronounced interference structure**.

## Conclusions

In this work, we presented and evaluated a reflection-geometry SHG fitting framework that combines the boundary-source treatment of Koopmans et al. [15] and Gielis et al. [19] with the distributed-bulk formalism of Koskinen et al. [21]. The detected field is modelled as the coherent sum of electric-dipole radiation from the surface, film bulk, and buried interface, including multiple reflections and propagation at both frequencies.

Measurements of a thick $Ta_2O_5$ film required both boundary and effective-bulk terms to reproduce all four polarisation channels within the assumed electric-dipole model. The synthetic-data analysis then established the conditions under which the fitted parameters can be interpreted: multiple polarisation configurations and accurate linear-optical inputs are required, while $\chi^{(2)}_{zzz}$ and responses with weak interference structure remain comparatively poorly constrained. These results also show that residual magnitude alone cannot be taken as proof of the correct SHG interpretation.

An important implication for future experiments is the fact that the SHG origin cannot be interpreted based only on the pP polarisation configuration, which is commonly used as the primary measurement scheme. Secondly, a fit of the SHG response of a layer with an optical thickness approaching the generating laser wavelength can also lead to good fit–data agreement without reflecting the actual physical reality. Nevertheless, in this case, the fit becomes unstable. Therefore, the issue can be identified by comparing several independently measured datasets and fitted $\chi^{(2)}$ values.

We also verified our findings experimentally. The co-deposited thick $Ta_2O_5$ films on N-BK7, N-SF10, and fused silica required the full model description to achieve good model-data agreement across the complete dataset. Moreover, the analysis showed comparatively similar fitted surface and effective-bulk responses but substrate-dependent buried-interface responses, consistent with the controlled change in substrate. Because the sample series intentionally modifies the buried interface while preserving the other source regions, it provides experimental support for the physical relevance of the source decomposition. Independent measurements at two positions on a 724 nm $Ta_2O_5$ film yielded closely similar parameters, unlike the analogous case of a 412 nm film, supporting our findings on experimental reproducibility depending on the layer thickness.

The method remains conditional on its assumptions. The fitted parameters are relative rather than absolutely calibrated; the bulk term is an effective descriptor rather than a microscopic identification; and electric-field-induced SHG, multipole terms, magnetic-dipole terms, and absorptive effects are excluded. The presented results should therefore be interpreted as validation of a source-resolved fitting strategy within a defined physical model, not as proof of a unique microscopic description. Absolute calibration, systematic uncertainty and correlation analysis, and controlled sample series linking the fitted responses to independently measured coating properties are the main priorities for future work.

With these extensions, the approach could support reference-oriented thin-film assessment. Coatings independently characterised with respect to adhesion, optical loss, mechanical stability, and other relevant material properties could define characteristic ranges of surface, bulk, and buried-interface responses. Newly deposited films could then be compared with these references to identify deviations and, where possible, localise them within the layered system.

## Acknowledgements

This work was co-funded by the project LasApp CZ.02.01.01/00/22_008/0004573 by the EU and the state budget of the Czech Republic; by the Student Grant Competition at the Technical University of Liberec (under the project No. SGS-2025-3577); and by the European Union (ERC, COINED, 101171288). Views and opinions expressed are, however, those of the authors only and do not necessarily reflect those of the European Union or the European Research Council Executive Agency. Neither the European Union nor the granting authority can be held responsible for them.

During the preparation of this manuscript, the authors used ChatGPT (GPT-6.1 Sol, OpenAI) to support language editing, literature search and summarisation, and manuscript formatting. The authors reviewed, validated, and critically assessed all AI-assisted outputs and take full responsibility for the content of the manuscript.

The authors declare no conflicts of interest.

## Data availability

The data that support the findings of this study, including the data underlying all figures, are openly available in Zenodo at https://doi.org/10.5281/zenodo.24093126.